# Positive magnetoconductance in SrVO$_3$ double quantum wells with a magnetic EuTiO$_3$ barrier

N. Takahara[1, 2], K. S. Takahashi[2, 3, *], Y. Tokura[1, 2, 4], and M. Kawasaki[1, 2]

[1] *Department of Applied Physics and Quantum-Phase Electronics Center (QPEC), University of Tokyo, Tokyo 113-8656, Japan*

[2] *RIKEN Center for Emergent Matter Science (CEMS), Wako 351-0198, Japan*

[3] *Department of Physics, Kanazawa University, Kanazawa 920-1192, Japan*

[4] *Tokyo College, University of Tokyo, Tokyo 113-8656, Japan*

\* Corresponding author: keitakahashi@se.kanazawa-u.ac.jp

**Abstract**

Controlling Mott insulator states has been a long-standing topic in condensed matter physics. Among various controlling parameters, two-dimensional (2D) confinement in epitaxial heterostructures has been demonstrated to convert the correlated metallic nature of SrVO$_3$ into a Mott insulator by reducing the quantum well thickness. Here, we fabricate double quantum well (DQW) structures of SrVO$_3$ with a magnetic barrier of EuTiO$_3$ to tune the hybridization of wave functions by a magnetic field. A significant positive magnetoconductance is observed for DQWs with barriers thinner than 2 nm, while DQWs with thicker barriers do not show such large positive magnetoconductance, instead behaving as a parallel circuit of two single QWs. The observed positive magnetoconductance is accounted in terms of enhanced hybridization of V 3*d* orbitals across the EuTiO$_3$ barrier under a magnetic field, where the barrier height is reduced by the Zeeman splitting of Ti 3*d* bands in forced ferromagnetic ordering of localized 4*f* electrons on Eu$^{2+}$ sites.

Strongly correlated electron systems have been investigated intensively for their intriguing physical properties that can be finely tuned not only by chemical substitution but also by external stimuli such as magnetic and electric fields as demonstrated for high temperature superconductivity in cuprates and colossal magnetoresistance in manganite [1]. Such external control pathways provide the opportunities for exploring potential devices, such as memories, switches, and sensors, which have promoted intensive studies employing epitaxial thin films and heterostructures, as well as for the development of elaborate physics models [2]. SrVO$_3$ (SVO) is a prototypical correlated electron metallic system that could be suitable testbed to explore the phase transitions by controlling the dimensionality in artificial quantum well (QW) structures. SVO is a perovskite transition metal oxide with V$^{4+}$ that has an electron configuration of $3d^1$ [3, 4]. Bulk crystals and thick films of SVO are a Pauli paramagnetic metal with low residual resistivity (of $\mu\Omega\ cm$ order) at low temperatures. However, LaVO$_3$, which contains V$^{3+}$ ions with a 3d$^2$ electron configuration, is a Mott insulator due to strong on-site Coulomb repulsion $U$. The solid solutions of La$_{1-x}$Sr$_x$VO$_3$ are Mott insulators for x < 0.18. Therefore, SrVO$_3$ is considered to be in an electronic state close to the Mott insulating one. According to Mott-Hubbard theory, the metal-insulator transition (MIT) can be controlled by varying the relative magnitudes of $U$ and bandwidth $W$, typically characterized by the ratio $U/W$. Using ultrathin films of SrVO$_3$, Yoshimatsu et al. observed a Mott-Hubbard gap via photoemission spectroscopy, which was attributed to the reduction of the effective bandwidth $W$ caused by a decrease in the effective coordination number at the interface and surface. We also reported the physical properties of ultra-thin SVO films sandwiched with insulating SrTiO$_3$ (STO) barrier layers to find that the system undergoes to a Mott insulating state for thicknesses less than a few unit cells (u.c.) [5]. Since the decrease in the layer thickness causes a reduction of kinetic energy gain along the out-of-plane direction, a two-dimensional (2D) Mott insulator state is thought to be realized as found in other oxides [6, 7]. We previously clarified that a phase transition takes place from an insulating state in 2D SVO films with thicknesses of 2 and 3 u.c. to a metallic states through an electron doping by the substitution of La for Sr at a narrow doping concentration around $x = 0.17$ [9]. This clearly indicates that 2D SVO could be considered as a typical Mott insulator [1]. If one can effectively control the dimensionality of the SVO QW by an external stimulus, a significant change in sheet conductance will be realized by a phase transition between Mott insulator and correlated metal. As one such method, we study tunnelling effect between two QWs in a double QW (DQW) structure as summarized in Fig. 1. Assuming a simple model, the tunneling transmittance $T$ across middle barrier layer can be approximately expressed as, $T = \exp\left(-\frac{b}{2\pi h}\sqrt{2mV}\right)$, where $b$ is the barrier thickness, $V$ the potential barrier height, $m$ the effective mass of electron, and $h$ Planck's constant, respectively [10]. Previously, DQW structures consisting of ultra-thin SVO (Mott insulator) and STO (band insulator) have been analyzed by angle-resolved photoemission spectroscopy (ARPES) to show the transition from a Mott insulator to a metal by reducing the barrier layer thickness [11]. In that work, heterostructures of SVO (2 u.c.)/STO ($n$ u.c.)/ SVO(6 u.c.)/STO(001) substrate were employed. Top SVO (2 u.c.) layer shows no metallic band dispersion with thick barriers ($n \geq 10$), while the metallic band dispersion appears in the top SVO layer by decreasing $n$ below 4. By the reduction of barrier layer thickness (corresponding to the reduction of $b$ in the above equation), the tunneling transmittance and hence the hybridization between the wave functions in top and bottom SVO layers were enhanced, resulting in the emergence of the metallic state in the top SVO layer [11].

Instead of changing the barrier thickness *b* as above, the phase transition from a Mott insulator to a metal can be designed if the potential barrier height *V* can be controlled by an external stimulus such as magnetic field as shown in Fig. 1F. In this study, we examine a magnetic insulator EuTiO$_3$ (ETO) as such a barrier layer. Since the crystal structure of ETO is a cubic perovskite, identical to that of SVO, the DQW can be epitaxially grown with atomically abrupt interfaces [12, 13, 14]. A typical structure of DQW is shown in Fig. 1A as a schematic and scanning transmission electron microscope (STEM) image. Here, two extreme cases are shown in Fig. 1B and 1C. In Fig. 1B, the DQW consists of two 3 u.c. QWs with a sufficiently thick barrier layer, maintaining an insulating state. In contrast, Figure 1C shows the case the 6 u.c. thick QW with metallic behavior, which can be regarded as a DQW composed of two 3 u.c. QWs without a barrier. Figure 1D shows the magnetization curve of ETO thin film with a thickness of 24 nm (60 u.c.) at 2 K. ETO exhibits G-type antiferromagnetic ordering below a Néel temperature $T_N$ of 5.5 K, while it can be easily forced to the ferromagnetic alignment by an external magnetic field as small as 3 T. In this forced ferromagnetic state, owing to the exchange coupling between Ti 3*d* and Eu 4*f* orbitals, the conduction Ti 3*d* bands of ETO exhibit a pronounced Zeeman splitting around 100 meV estimated by ab-initio calculations [15], as schematically represented in Fig. 1F. In fact, this Zeeman splitting was elucidated experimentally in our previous study in the analysis of the Shubnikov-de Haas oscillations observed in high electron mobility films of La doped ETO [15]. We expect that the barrier height should be tuned by opening the Zeeman splitting gap as depicted in the right panel of Fig. 1F. By applying magnetic field, the degenerated Ti $t_{2g}$ bands of ETO are split into up and down spin bands, resulting in a decrease in the tunneling barrier height for ETO and facilitating the hybridization of wave functions between the two SVO layers. Thus, we expect that a magnetic field can change the 'total' sheet conductance of DQW parallel to the tunnel barrier. The measurement configuration for this possible magnetoconductance is shown in Fig. 1E.

In this paper, we discuss the transport properties of a series of DQWs with SVO wells and either middle barrier layers of magnetic ETO or nonmagnetic STO. When the barrier layer thickness is reduced below 4 u.c., a significant increase was observed in sheet conductance for both barrier layers, indicating the contribution of the tunneling effect. Below this critical thickness, the ETO DQW exhibits a positive magnetoconductance, attributed to the reduction in barrier height caused by the Zeeman splitting in the ETO $t_{2g}$ band, in contrast to the negative magnetoconductance observed in the STO DQW.

The DQWs were grown on $(LaAlO_3)_{0.3}(Sr_2AlTaO_6)_{0.7}$ (LSAT) (001) substrates by using a gas-source MBE [16, 17, 18], employing titanium tetra-isopropoxide and vanadium tri-isopropoxide and metal sources of Sr and Eu without supplying any oxidizing gas as described in our previous reports [5, 9, 15]. The sample structure is STO (5 u.c.)/SVO (3 u.c.)/ETO (STO) ($n$ u.c.)/SVO (3 u.c.)/STO (5 u.c.), abbreviated as SV/ET (ST) ($n$ u.c.)/SV in the followings. The thicknesses of layers are controlled by the deposition time. Figure 1A shows a cross-sectional image for an SV/ET (5 u.c.)/SV taken by a high-angle annular dark-field scanning transmission electron microscope (HAADF-STEM) and the elemental maps, exhibiting a well-ordered atomic alignment with high crystallinity and indiscernible defect. The elemental maps are shown for the $L$ edges of Sr and Eu and the $K$ edges of Ti and V obtained through energy dispersive X-ray spectrometry (EDX) for the same area as in Fig. 1A, revealing the abrupt interfaces.

We now discuss the transport properties for the DQWs measured by the four-probe method as shown in Fig. 1E. Figures 2A and B show the temperature dependence of sheet conductance $G$ for SV/ET($n$ u.c.)/SV and SV/ST($n$ u.c.)/SV, respectively. The data for SVO (6 u.c.) single QW denoted as $n = 0$ and twice the value for SVO (3 u.c.) single QW as $n = \infty$ are also shown for references of extreme cases of the DQWs. Here, SVO with a thickness of 3 u.c.is a Mott insulator, while that of 6 u.c. is a metal. This leads to a two orders of magnitude difference in conductance between the $n = 0$ and $n = \infty$, despite the total thickness of SVO being the same. The thickness dependence of $G$ at 2 K is shown in Fig. 2C. In the course of decreasing the barrier layer thickness $n$ in the DQWs from 10 to 1, the conductance $G$ for both SV/ET/SV and SV/ST/SV increases sharply below $n = 3$ as indicated by the purple region, while $G$ almost remains nearly constant for DQWs from $n = 5$ to 10. Significant enhancement in $G$ is plausibly due to the hybridization of 3$d$ electron wave functions in two QWs facilitated by the tunneling effect across the barrier layer, resulting in a phase transition into a metallic state for the samples with $n <$ 5. This behavior is drawn as schematics of left and right panels in Fig. 2C. The critical thickness of $n = 3$ is comparable to that observed in the previous study by ARPES measurements [11]. There is small but systematic difference in $G$, being smaller for SV/ET/SV structures than SV/ST/SV structures with the same barrier thickness. This is presumably due to the lower effective barrier height for the STO layer, possibly due to oxygen deficiency that induces a shallower Fermi energy in STO compared to ETO.

Figure 3 shows magnetic field dependence of magnetoconductance (MC) at 2 K defined as $\Delta G = G(B) − G(0)$ for SV/ET($n$ u.c.)/SV and SV/ST($n$ u.c.)/SV. To clarify the behaviors of the MC, the barrier thickness dependence of $\Delta G$ at $B = 3$ T is plotted in Fig. 3E for both DQWs. Generally, a negative MC emerges in a paramagnetic metal with $B^2$ dependence, which can be attributed to a Lorentz motion of electrons [19]. In fact, the sign of the MCs of non-magnetic SV/ST/SV DQWs is found to be negative as seen in Fig.

3C, D, and E. The negative MC becomes pronounced in the metallic DQWs of the tunneling regime ($n = 1 – 3$), although the magnetic field dependence deviates from the expected $B^2$ function. In contrast to the SV/ST/SV, a prominent feature in magnetic SV/ET/SV is a large positive MC observed in the tunnel regime ($n = 2 – 3$). Furthermore, a very small but distinctly positive MC is also discernible in the thick barrier regime ($n = 5 – 10$) as well, where the two QWs are decoupled due to the low tunnel transmittance. We speculate that the origins of the positive MC observed for SV/ET/SV in the two regimes are different. For the thin ETO barrier regime, as illustrated in the left panels of Fig. 3E, the positive magnetoconductance (MC) can be attributed to the magnetically enhanced tunneling through the ETO layer as we have expected. On the other hand, the small positive MC observed for the thick ETO barrier regime is likely to originate from an interface spin-scattering effect. Such magnetic contributions from the ETO layer can be observed in the temperature dependence of the positive magnetoconductance. As shown in Supplemental Material Fig. S3, the magnetoconductance gradually diminishes with increasing temperature, following a similar temperature dependence as the magnetization.

One might consider that the positive magnetoconductance is attributable to a weak localization. In fact, a positive magnetoconductance have been observed, especially at high magnetic fields, in SVO single QWs with a certain thickness, implying the contribution of weak localization effect as shown in supplementary Fig. S1. So far, we have been unable to convincingly explain such non-monotonic magnetoconductance observed in the single QWs. However, the DQWs with a 2 or 3 u.c. thick ETO barrier layer exhibit a sharp and pronounced positive magnetoconductance at low magnetic fields as shown by red curves in Fig. S1, likely indicating a relation with the magnetism of ETO.

The conduction electrons in SVO QWs adjacent to ETO are slightly spin-polarized through the exchange interaction with the aligned Eu spins under applied magnetic field, causing the small positive MC as illustrated in the right panels of Fig. 3E [20, 21, 22]. It is notable that the SVO DQW with the thinnest ETO barrier ($n = 1$) exhibits negative magnetoconductance. Although such behavior might suggest an instability of the magnetic order in a single monolayer of Eu spins, further experimental evidence is needed for clarification. The positive magnetoconductance for $n = 2$ is smaller than that for $n = 3$, and the corresponding slope, particularly in the low magnetic field region, is noticeably more gradual. We suspect that the spin structure and each Eu spin moment in the ETO layer with a thickness of $n = 2$ are more unstable than those in the layer with $n = 3$. Therefore, the positive magnetoconductance caused by the reduction in tunnelling barrier height is less effective in the $n = 2$ DQW than in the $n = 3$.

We now compare the magnetic field dependences of the MC (Fig. 3A and B) with the magnetization of a thick ETO film (Fig. 1D). One expects a scaling relation between the positive MC and the magnetization of ETO barrier layer, but they do not exhibit similarity. Namely, the MCs keep increasing without any sign of saturation up to 4 T, while the magnetization saturates at 3 T. Interestingly, the magnetic field dependence of MC is similar between the tunnel regime ($n = 2$ and 3 in Fig. 3A) and interface one ($n = 5, 8, 10$ in Fig. 3B). Therefore, one possible scenario is that the magnetization curve of thin ETO below 10 u.c. is slightly modified from that of the thick ETO films as Fig. 1D and does not saturate even at 4 T, which contributes to the continuously increasing MC without saturation. Unfortunately, it is difficult to accurately measure the magnetization curve of such ultra-thin films due to the significant diamagnetic contribution from the substrate in the SQUID signal. Elucidating the relationship between the MC and magnetization is left for future study. We note that a positive magnetoconductance is also observed under the in-plane magnetic field under in-plane magnetic fields, Although the magnitude of the positive magnetoconductance is smaller than that under the out-of-plane magnetic field, the magnetic field dependence of the positive magnetoconductance is significantly similar. Since the magnetization of EuTiO$_3$ is isotropic, the only anisotropy present in thin films arises from the shape anisotropy due to the demagnetizing effect. Therefore, such directional differences in magnetoconductance are to be addressed in future studies. Using the equation of $T = \exp\left(-\frac{b}{2\pi h}\sqrt{2mV}\right)$, we can deduce the modulation of the barrier height caused by the Zeeman splitting in the ETO layer. A quantitative analysis, based on several assumptions, is provided in the Supplemental Material.

We have studied transport property of SVO double quantum well structures with magnetic ETO and non-magnetic STO barriers. By using gas-source molecular beam epitaxy, high crystalline films were obtained, and barrier layer thickness was controlled systematically at the unit cell scale. The sheet conductance exhibits a sharp increment below *n* = 4, being indicative of the hybridization of wave functions across the barrier layer due to the tunneling effect. In contrast to the negative magnetoconductance observed in SVO/STO/SVO structures, positive magnetoconductance is observed in SVO/ETO/SVO structures. This is ascribed to the forced ferromagnetic ordering in ETO, which contributes to both the suppression of magnetic scattering at interface and the reduction of tunneling barrier height due to Zeeman splitting. These findings will give a new insight to further research on the quantum size effect of Mott transition in artificially constructed oxide heterostructures.

ACKNOWLEDGMENTS

This work was partly supported by JSPS KAKENHI (22H04958 and 23H01857) and Mitsubishi Foundation.

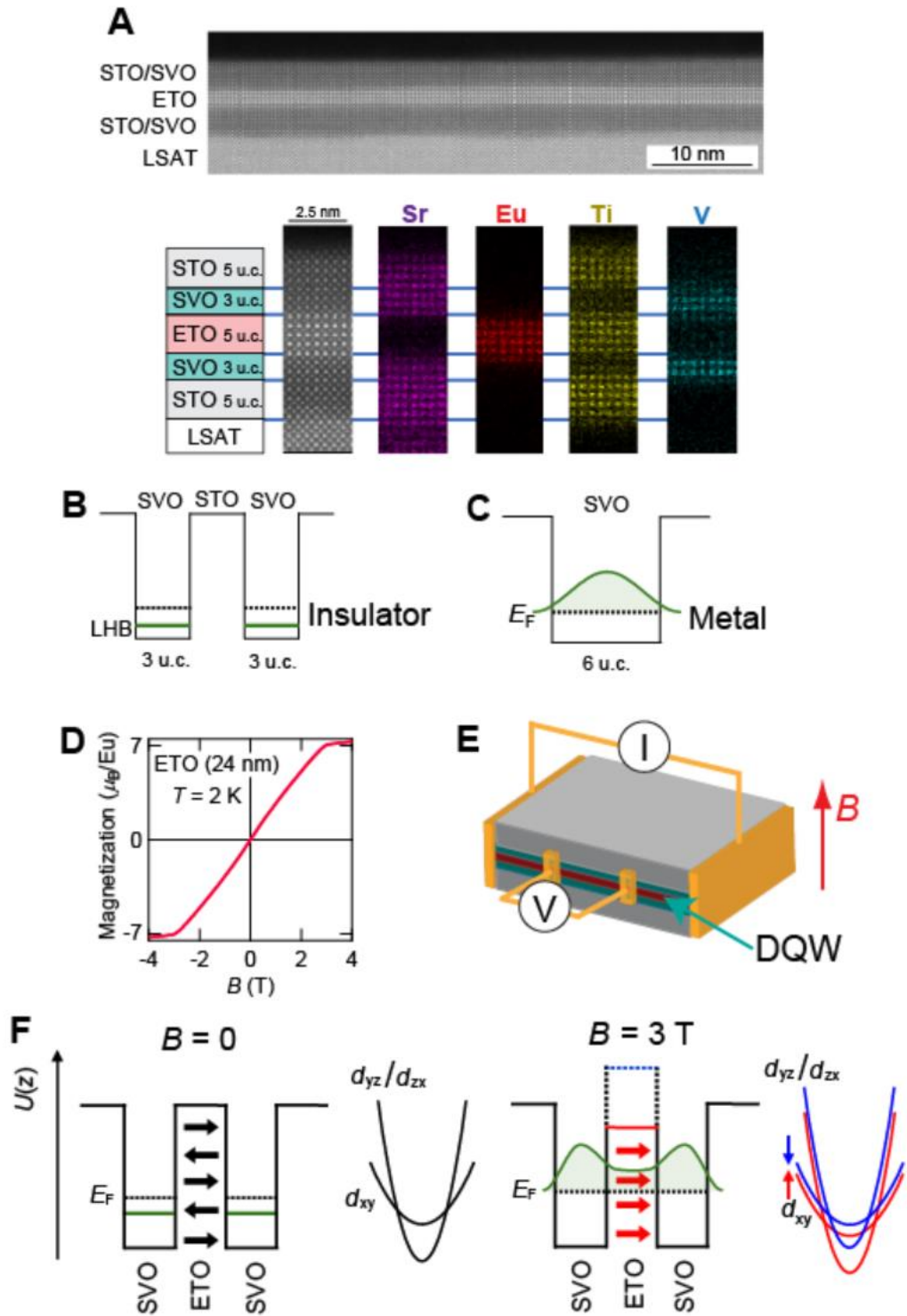

Fig. 1 (A) A cross-sectional image taken with a high-angle annular dark-field scanning transmission electron microscope (HAADF-STEM) viewed along the [100] axis of a double quantum well structure consisted of SrTiO$_3$ (STO) (5 u.c.), SrVO$_3$(SVO) (3 u.c.)

and EuTiO$_3$ (ETO) ($n$ = 5 u.c.) grown on (LaAlO$_3$)$_{0.3}$(SrAl$_{0.5}$Ta$_{0.5}$O$_3$)$_{0.7}$ (LSAT) substrate. The elemental maps for Sr, Eu, Ti, and V deduced from an energy dispersive x-ray spectrometry (EDX), where $L$ ($K$) edge was taken for Sr and Eu (Ti and V). Schematics for (B) double quantum well of two 3 u.c. SVO layers in $d^1$ Mott insulator state and (C) thick (6 u.c.) single quantum well in $d^1$ metallic state. (D) Magnetization curve at 2 K for a 24 nm thick ETO film on LSAT substrate. (E) A sketch of a four probe measurement for the magnetoconductance parallel to the DQW. (F) Schematics of electronic states in the double quantum well with ETO barrier layer and the Ti $t_{2g}$ band dispersion along $k_z$ direction for the ETO under $B$ = 0 T (left panel) and 3 T (right panel). When the magnetization of ETO is saturated at $B$ = 3 T, the Ti $t_{2g}$ bands are split by Zeeman effect (red (blue) curves correspond to up (down) spin bands). We expect that such band splitting reduces the potential of tunneling barrier in ETO, promoting a hybridization between the electron wave functions in two quantum wells.

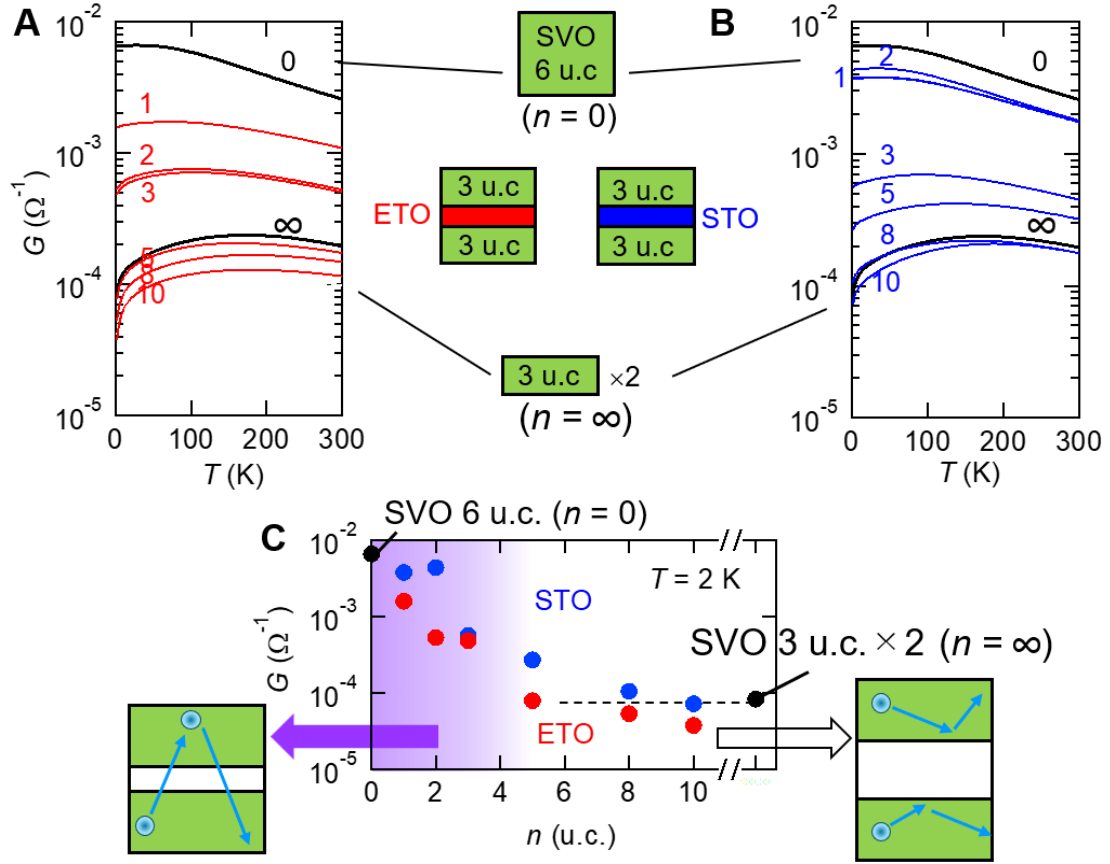

Fig. 2 Temperature dependence of sheet conductance $G$ for (A) STO/SVO/ETO/SVO/STO (red) and (B) STO/SVO/STO/SVO/STO (blue) double quantum well structures with respective barrier layers of ETO and STO with thicknesses $n$ = 1-10 u.c.. Black curves are the sheet conductance for SVO (6 u.c.) single layer film taken as $n$ = 0 case (upper) and doubled value of sheet conductance for SVO (3 u.c.) single layer film taken as $n = \infty$ case (lower). (C) Sheet conductance at $T$ = 2 K as a function of barrier layer thickness $n$ for STO/SVO/ETO/SVO/STO (red) and STO/SVO/STO/SVO/STO (blue). The purple shadow indicates the range of barrier thicknesses where the conductance enhancement takes place due to the tunneling-induced hybridization effect.

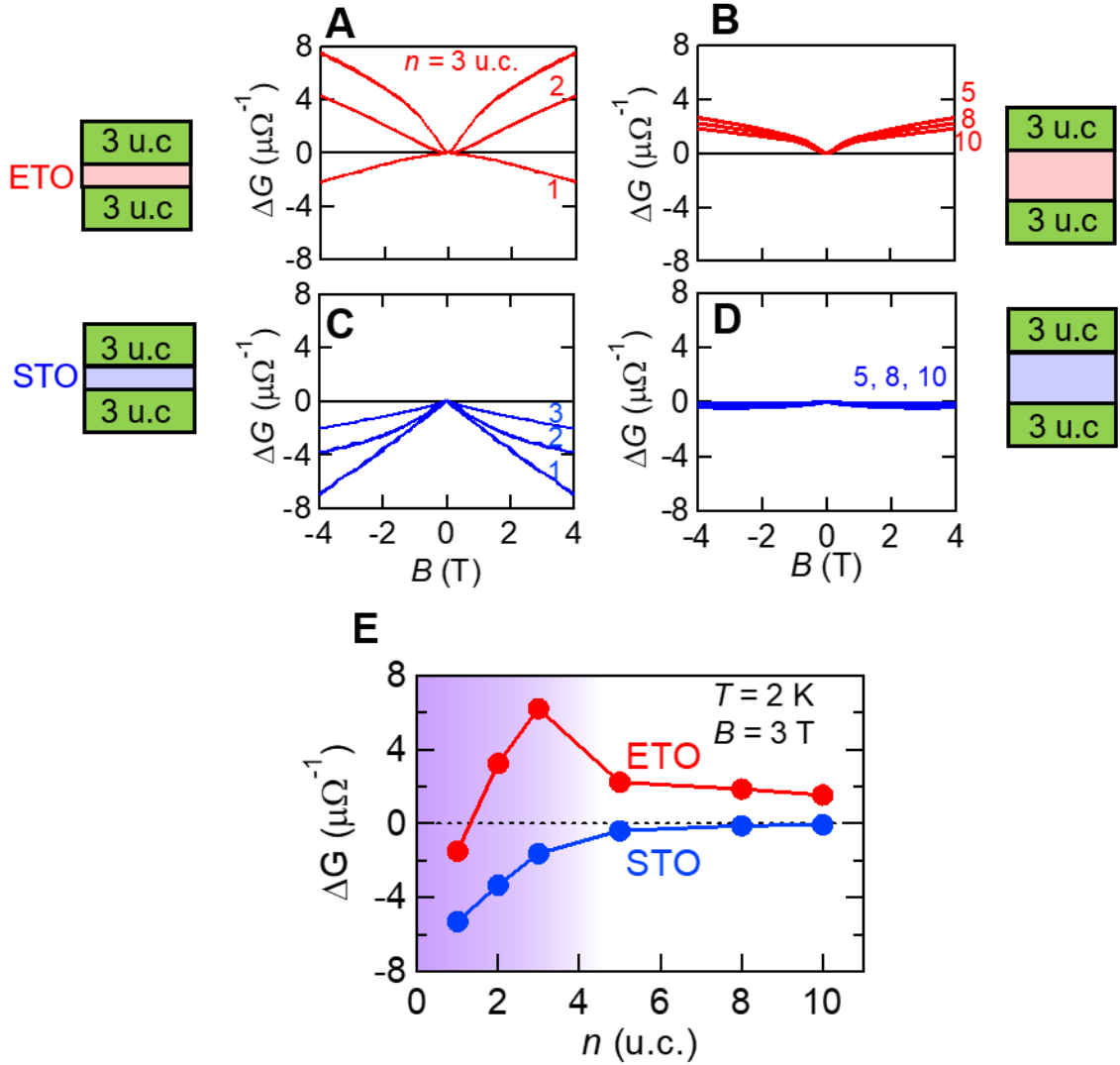

Fig. 3 (A-D) Magnetic field dependence of magnetoconductance $\Delta G = G(B) - G(0)$ at $T = 2$ K for (A, B) STO/SVO/ETO/SVO/STO and (C, D) STO/SVO/STO/SVO/STO double quantum well structures with respective barrier layers of ETO and STO with thicknesses of $n = 1\text{-}10$ u.c.. (E) Magnetoconductance at $T = 2$ K as a function of barrier layer thickness $n$ for STO/SVO/ETO/SVO/STO (red) and STO/SVO/STO/SVO/STO (blue) at $B = 3$ T. Purple region indicates the barrier thickness where conductance enhancement occurs for ETO barrier samples due to the tunneling-induced hybridization effect.